\documentclass[fleqn,usenatbib,useAMS]{mnras}

\newcommand{\comment}[1]{}

\newcommand{\changed}[1]{{#1}}

\usepackage{graphicx}	
\usepackage{amsmath}	
\usepackage{amssymb}	
\usepackage{multicol}        
\usepackage{bm}		

\usepackage[T1]{fontenc}
\usepackage{ae,aecompl}

\title[SCIDAR turbulence velocity profiles]{\changed{Turbulence velocity profiling for high sensitivity and vertical-resolution atmospheric characterisation with Stereo-SCIDAR}}

\author[J. Osborn]{J. Osborn\thanks{Contact e-mail: \href{mailto:james.osborn@durham.ac.uk}{james.osborn@durham.ac.uk}}, T. Butterley, M.~J. Townson, A.~P. Reeves, T.~J. Morris and R.~W. Wilson
\\
$^{1}$Centre for Advanced Instrumentation, Department of Physics, Durham University, South Road, Durham, DH1 3LE, UK}

\date{\today}

\pubyear{2016}

\begin{document}
\label{firstpage}
\pagerange{\pageref{firstpage}--\pageref{lastpage}}
\maketitle

\begin{abstract}
As telescopes become larger, into the era of $\sim$40~m Extremely Large Telescopes, the high resolution vertical profile of the optical turbulence strength is critical for the validation, optimisation and operation of optical systems. The velocity of atmospheric optical turbulence is an important parameter for several applications including astronomical adaptive optics systems. Here, we compare the vertical profile of the velocity of the atmospheric wind above La Palma by means of a comparison of Stereo-SCIDAR with the Global Forecast System models and nearby balloon borne radiosondes. We use these data to validate the automated optical turbulence velocity identification from the Stereo-SCIDAR instrument mounted on the 2.5~m Isaac Newton Telescope, La Palma. By comparing these data we infer that the turbulence velocity and the wind velocity are consistent and that the automated turbulence velocity identification of the Stereo-SCIDAR is precise. The turbulence velocities can be used to increase the sensitivity of the turbulence strength profiles, as weaker \changed{turbulence} that may be misinterpreted as noise can be detected with a velocity vector. The turbulence velocities can also be used to increase the altitude resolution of a detected layer, as the altitude of the velocity vectors can be identified to a greater precision than the native resolution of the system. \changed{We also show examples of complex velocity structure within a turbulent layer caused by wind shear at the interface of atmospheric zones.}
\end{abstract}

\begin{keywords}
atmospheric effects, instrumentation: adaptive optics, methods: data analysis, methods: statistical, site testing
\end{keywords}




\section{Introduction}
As astronomical adaptive optics (AO) systems become more sophisticated, a detailed knowledge of the altitude profile of optical turbulence, measured as the refractive index structure constant, $C_n^2\left(h\right)$, and velocity, is increasingly important. Detailed characterisation of the atmospheric turbulence profile permits realistic modelling, performance prediction, and real time validation and optimisation of AO instruments. 

The large number of wavefront sensor subapertures across the AO system pupil combined with the wide field of view means that the next generation of 30 to 40~m class Extremely Large Telescopes (ELTs) will be significantly more sensitive to variations in the optical turbulence profile than existing 8~m class systems. These telescopes will be sensitive to variations in turbulence altitude of the order of 100 to 500~m \citep{Neichel2008, Basden2010, Vidal2010, Masciadri2013a, Gendron2014}. This is currently a challenge fot optical turbulence profiling instrumentation.

The three most prevalent optical profiling techniques in operation today are MASS (Multi Aperture Scintillation System, \citealp{Tokovinin07}), SCIDAR (SCIntillation Detection And Ranging, \citealp{Vernin73}) and SLODAR (SLOpe Detection And Ranging, \citealp{Wilson2002}). MASS is not intended as a high vertical-resolution technique. It has a limited logarithmic vertical resolution and the high altitude response is very broad \citep{Tokovinin07}. Both SLODAR and SCIDAR are triangulation techniques in which the atmospheric turbulence profile is recovered from either the correlation of wavefront slopes in the case of SLODAR, or scintillation intensity patterns in the case of SCIDAR, for two target stars with a known angular separation. Generalised-SCIDAR \citep{Fuchs94} is a development of the SCIDAR technique where the detectors are conjugate below the ground level in order to make the instrument sensitive to turbulence at the ground.
For SLODAR the resolution is limited by the double star separation and the size of the wavefront sensor subapertures. The subapertures must be sufficiently large ($> 5-10$ cm, depending on stellar magnitude) to provide adequate signal for centroid measurements. To achieve a vertical turbulence profile up to 20~km with 200~m altitude resolution, the SLODAR would require 100 subapertures, with 0.08~m suparpertures this would require an 8~m telescope.

The theoretical resolution for Generalised-SCIDAR varies with the Fresnel zone size for a given altitude of the \changed{turbulence} and is given by \citep{Prieur01},
\begin{equation}
	\delta h\left(z\right) = 0.78\frac{\sqrt{\lambda z}}{\theta},  
	\label{eq:dh}
\end{equation}
where $\lambda$ is the wavelength, $\theta$ is the separation of the double star and \changed{$z$ is the propagation distance from the turbulence to the detection plane. $z$ depends on the airmass of the observation ($\sec{(\theta_z)}$, where $\theta_z$, is the zenith angle), the conjugate altitude of the detector plane, $z_{\mathrm{conj}}$, and the altitude of the turbulent layer, $h$, and is given by, $z=| h\sec{(\theta_z)} +z_{\mathrm{conj}}|$}. For larger propagation distances the spatial scale of the intensity speckle patterns is larger, reducing the altitude resolution. Therefore the native altitude resolution of Generalised-SCIDAR is altitude dependent with a significantly reduced resolution for high-altitude turbulence. As with SLODAR, in order to achieve 200~m altitude resolution to 20~km with Generalised-SCIDAR an 8~m telescope is required \citep{Masciadri2013a}.

\cite{Egner2007} proposed a way to increase the altitude resolution of Generalised-SCIDAR based on simultaneous turbulence velocity profiles. To implement this High-Vertical Resolution (HVR) mode the profiler needs to be able to measure the vertical profile of the turbulence velocity in addition to the strength. The approach taken is to track the position of the covariance peaks in the spatio-temporal cross-covariance function. Using the frame rate and pixel size the turbulence velocity can be estimated. Differential velocity vectors within a vertical resolution element signify different turbulent layers and can provide a vertical resolution better than the native resolution of the instrument. Although turbulence velocity profiling has been demonstrated with SLODAR (for example \citealp{Cortes2012,Guesalaga2014,Gilles2013} and Generalised-SCIDAR (for example, \citealp{Prieur04,Garcia_lorenzo06}) an automated approach is difficult. 

This improved altitude resolution, enabled by the high-vertical resolution mode, is currently of critical importance for studies for the next generation of ELTs. 

For SLODAR, the number of subapertures in existing systems is generally low, making it difficult to separate the covariance peaks even in the temporal dimension. For Generalised-SCIDAR there are three covariance peaks for each turbulent layer. This excess of signals is difficult for automated systems to track.

Stereo-SCIDAR \citep{Shepherd13,Osborn13} is a Generalised-SCIDAR instrument which is designed to measure the vertical profile of optical turbulence strength and velocity in the full atmosphere. In contrast to most SCIDAR instruments, Stereo-SCIDAR uses two cameras, one to image the defocussed pupil of each of the two target stars. By doing this {the intensity of the images can be normalised independently, and hence }Stereo-SCIDAR has increased sensitivity to weaker \changed{turbulence} and can operate with a larger difference in brightness of the target stars. In addition, the spatial cross-covariance function has only one covariance peak per turbulent layer. This lends itself to automated turbulence velocity identification over the full atmosphere \citep{Osborn2015}. 

Stereo-SCIDAR can use the automated velocity profiles with the HVR technique of \citep{Egner2007} to measure the optical turbulence with ELT-scale altitude resolution (100 - 500~m). In addition, the turbulence velocity identification can be used to confirm the existence of weak turbulent layers close to the noise floor of the instrument. If a covariance peak is seen to move in the spatio-temporal covariance function then we can confirm that it is real and not simply noise, which would behave differently. \changed{This turbulence} could easily be ignored in the data analysis but for applications where high sensitivity is required, they could be important. However, before this data can be used in performance simulations for future ELT instrumentation the automated turbulence velocity identification must be validated.


Here, we compare wind and turbulence velocity profiles from three sources: 
\begin{itemize}
\item{physical tracking measurements of balloon borne radiosondes via a Global Positioning System and radiotheodolite.}
\item{computer model, Global Forecast System (GFS) meteorological forecasts provided by the National Oceanic and Atmospheric Administration \citep{NOAA2015}}
\item{optical remote sensing, Stereo-SCIDAR automated wind velocity detection algorithm.}
\end{itemize}

These data sources were chosen for our comparison as both the measured radiosonde and modelled GFS wind velocities have been shown to provide precise estimates (for example, \citealp{Sarazin2011,Vernin1997,Garcia_lorenzo06,Avila2006}).

With this cross-validation we show that the turbulence and the wind velocity are highly correlated, simultaneously showing that the Stereo-SCIDAR turbulence velocity is precise and that the numerical models of the wind velocity can be used as an estimator for the turbulence velocity.

In addition to improving the altitude resolution of optical profilers the wind velocity vertical profile is an important parameter for astronomical Adaptive Optics (AO) applications.

The wind velocity determines how quickly an AO system must be updated. For AO systems this can be parametrised by the coherence time, a measure of how long the turbulence can be assumed to be coherent, \citep{Greenwood1977}. The coherence time defines the update rate at which an AO system must function at in order to minimise residual wavefront errors due to the temporal lag between the wavefront measurements and correction on the deformable mirror (DM). Higher wind speeds mean that the coherence time will be shorter and the AO system update rate will have to be faster. However, the optical turbulence strength and velocity are not constant in altitude and so to calculate the coherence time we need to know the speed of the turbulence at the altitude of the turbulence. Therefore, the vertical profile of the turbulence velocity as well as the turbulence strength is required to calculate the coherence time, an important parameter for both the design and real-time operation of AO systems.

Sophisticated AO controllers that make use of the wind velocity profile can also be used to improve the performance of AO systems. One such technique is Linear Quadratic Gaussian (LQG) \citep{Paschall1993,Kulcsar2006}. In this case, a spatial and temporal model is used to reconstruct the wavefront. LQG has been tested on-sky and proven to provide superior AO reconstruction and control over conventional integrators \citep{Sivo2014,Sivo2014b}. 
In the case of predictive distributed Kalman filters (DKF), \cite{Gilles2013} show that the wind profile estimation must be accurate to better than 20\%.

\changed{
The wind velocity profiling algorithm outlined above assume "frozen flow", i.e. that the wavefront does not evolve as it traverses the telescope field of view. However, the interface of two zones of atmosphere with different wind velocities can induce a velocity dispersion in the turbulence.  Although the optical turbulence is a passive tracer of the bulk motion of the air, a gradient of velocity between the top and the bottom of the turbulent zone can exist. This may not be seen in numerical models and probes (such as weather balloon tracking) due to the discrete altitude sampling. Optical systems, such as AO, are sensitive to the velocity of the optical turbulence (refractive index variations) and not necessarily the bulk velocity of the wind. It is only by comparing the numerical models with turbulence velocities determined by optical remote sensing means that we can understand the differences, if any, between wind velocity, turbulence velocity and gusting. The Stereo-SCIDAR has sufficient resolution to be able to resolve this and we show examples of such velocity dispersion.
}

In section~\ref{sect:inst} we describe the instrumentation used in this study and in section~\ref{sect:comp} we show comparisons of the wind and turbulence velocity measurements. Section~\ref{sect:conc} describes the conclusions from this study.

\section{Instrumentation}
\label{sect:inst}
\subsection{Stereo-SCIDAR}
\begin{figure}
\centering
\includegraphics[width=0.3\textwidth]{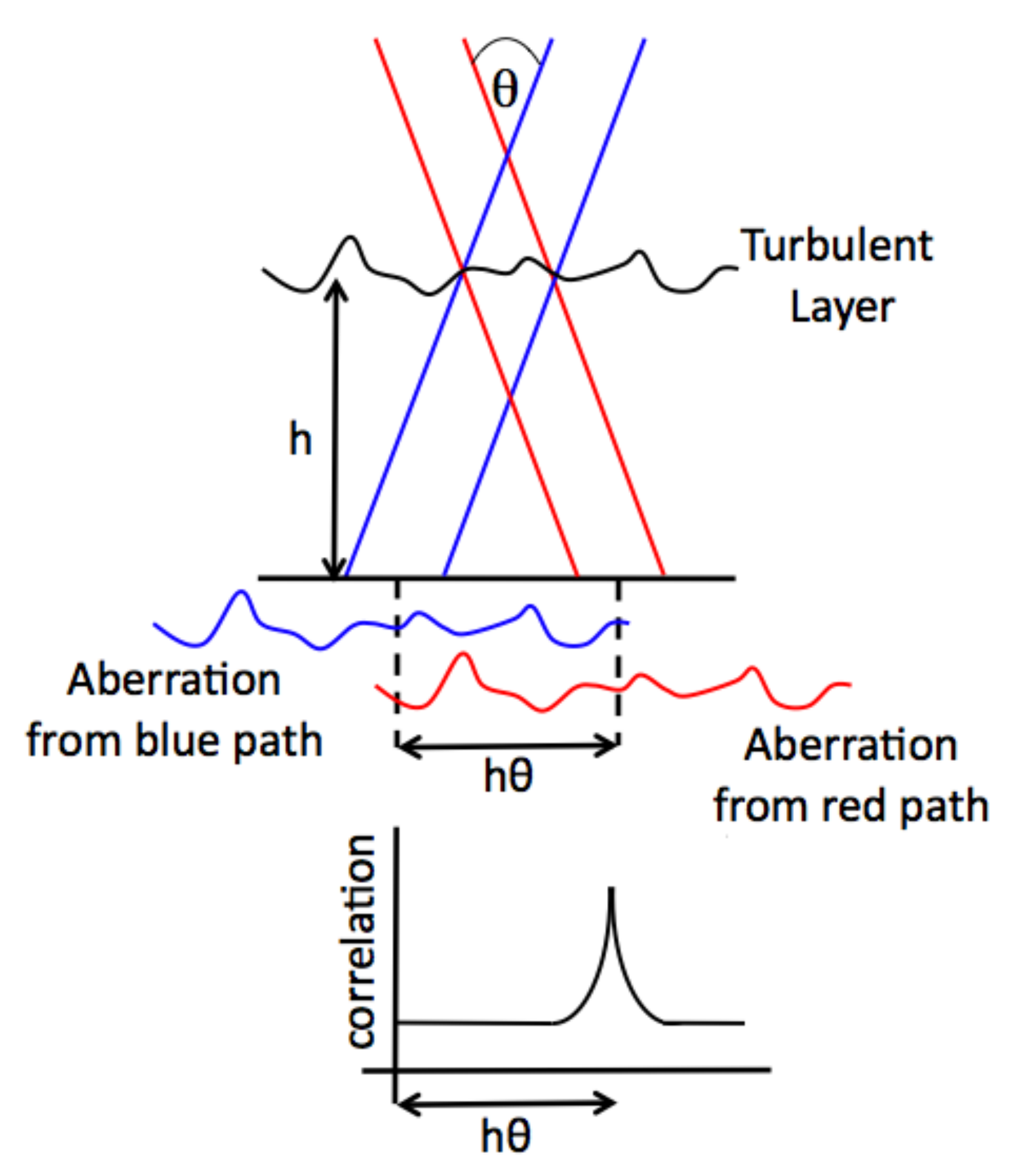}
\caption{If a turbulent layer at height, $h$, is illuminated by two stars of angular separation, $\theta$, then two copies of the aberration will be made on the ground separated by a distance $h\theta$. By cross correlating either the centroid positions from a Shack-Hartmann wavefront sensor (SLODAR) or the intensity patterns (SCIDAR) we can triangulate the height of the turbulent layer and the amplitude of the correlation peak corresponds to the strength of the layer.}
\label{fig:triangulation}
\end{figure}

SCIDAR is a technique to profile the vertical distribution of atmospheric turbulence by triangulation. The distance to a turbulent layer is estimated by measuring the spatial displacement of the scintillation patterns from two target stars. The strength of the turbulence is related to the magnitude of the covariance peak (figure~\ref{fig:triangulation}). The technique can be extended by optically conjugating the detectors to an altitude below the telescope and in this way the Generalised-SCIDAR can also be used to measure the turbulence at the ground \citep{Fuchs98}.

Stereo-SCIDAR is a stereoscopic version of Generalised-SCIDAR, in which a separate detector is used to image each star. The reader is referred to \citep{Shepherd13} for a full description of Stereo-SCIDAR, including how it works and the opto-mechanical design.

\begin{figure}
\centering
\hspace*{-10mm}\includegraphics[width=0.5\textwidth,trim={0cm 1cm 1.5cm 1cm}]{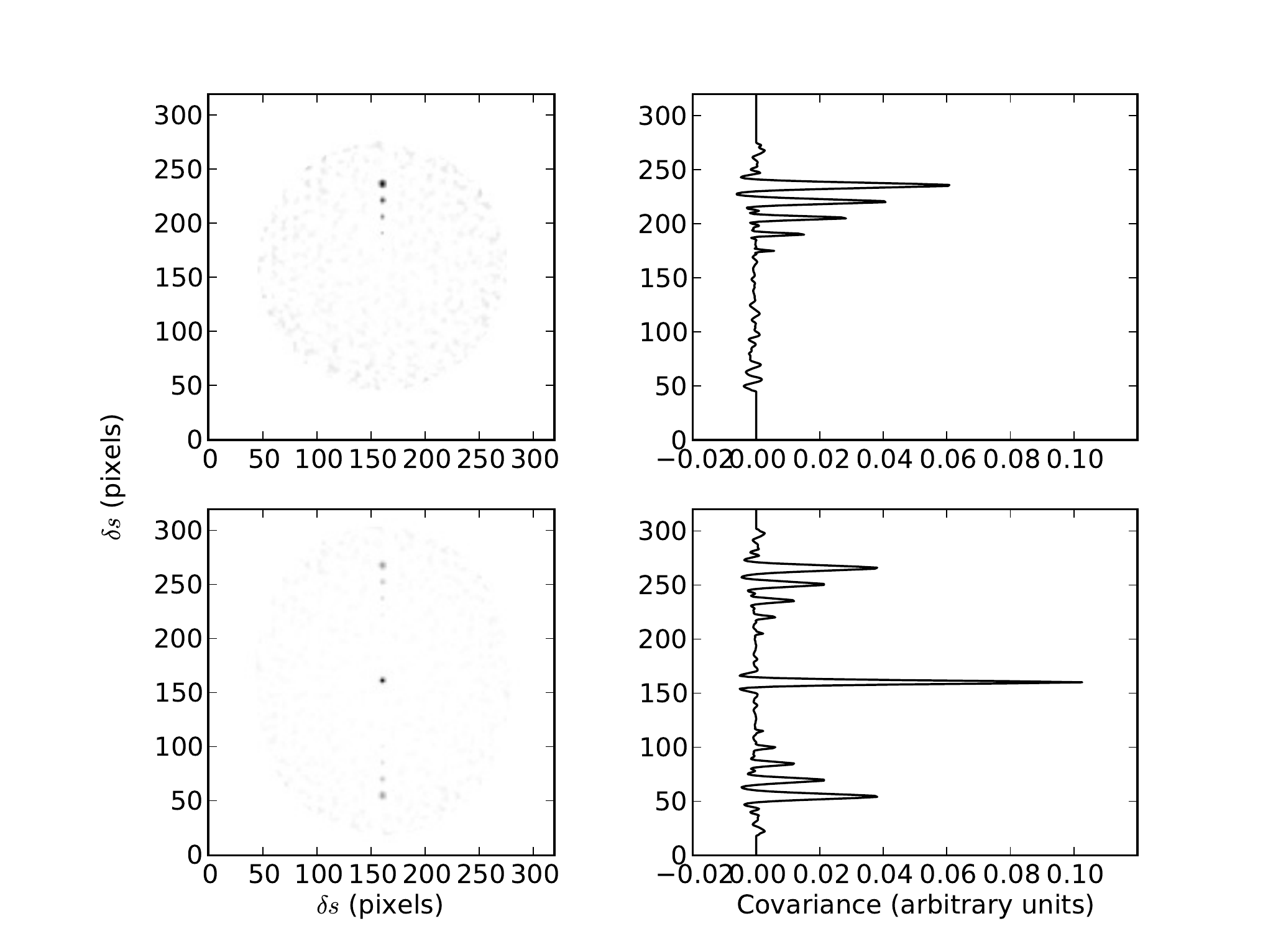}
\caption{Simulated 2D covariance functions (left column) and vertical cut through (right column) for Stereo-SCIDAR (top row) and conventional generalised-SCIDAR (bottom row). In this case the atmospheric simulation contained six equal strength turbulent layers at 2 km spacing between 0 and 10 km, inclusive.}
\label{fig:xcov}
\end{figure}

By separating the scintillation patterns onto separate detectors instead of overlapping them on a single camera (as with traditional SCIDAR instruments) we reduce the noise in the profile estimation. This is because, in conventional SCIDAR instruments, the intensity speckles lose contrast in the overlapping patterns, reducing the visibility of the covariance peaks (figure~\ref{fig:xcov}). A vertical cut through each covariance function is shown on the right. $\delta s$ is the position in the covariance function. We see that for single camera SCIDAR we have two sets of spatially separated peaks and one set of overlapping peaks at the centre. For Stereo-SCIDAR we only have one set. Both plots have the same contrast scale, the correlation peaks for Stereo-SCIDAR are larger in magnitude than that of single camera Generalised-SCIDAR.

Stereo-SCIDAR was designed, built and operated as part of the CANARY project, a demonstrator for ELT-scale AO technologies on the 4.2~m William Herschel Telescope (WHT), La Palma \citep{Morris2014,Martin2014}. For this project the Stereo-SCIDAR was operated on the 2.5~m Isaac Newton Telescope (INT), approximately 400~m from the WHT \citep{Osborn2015b}.

The Stereo-SCIDAR on the INT has 100 pixels across the 2.54~m re-imaged pupil, resulting in 2.54~cm effective pixel size. The frame rate is $\sim$100~Hz and the exposure time is 2~ms\changed{ to ensure that enough flux is received for full time coverage i.e. that a useable target is always visible.} The electron multiplication gain on the detectors is chosen such that the images do not saturate the detectors and usually operate with a peak intensity at $\sim$80\% of the maximum saturation level. A dichroic beam splitter with a cutoff at 615~nm is used to divert the red light onto a SLODAR channel for a dual SLODAR/SCIDAR experiment \citep{Butterley2015}. The shorter wavelength light goes to the Stereo-SCIDAR instrument.

\subsubsection{Measuring wind velocity with Stereo-SCIDAR}

\begin{figure*}
\centering
$\begin{array}{ccccc}
	\hspace{-4mm}\includegraphics[width=0.22\textwidth]{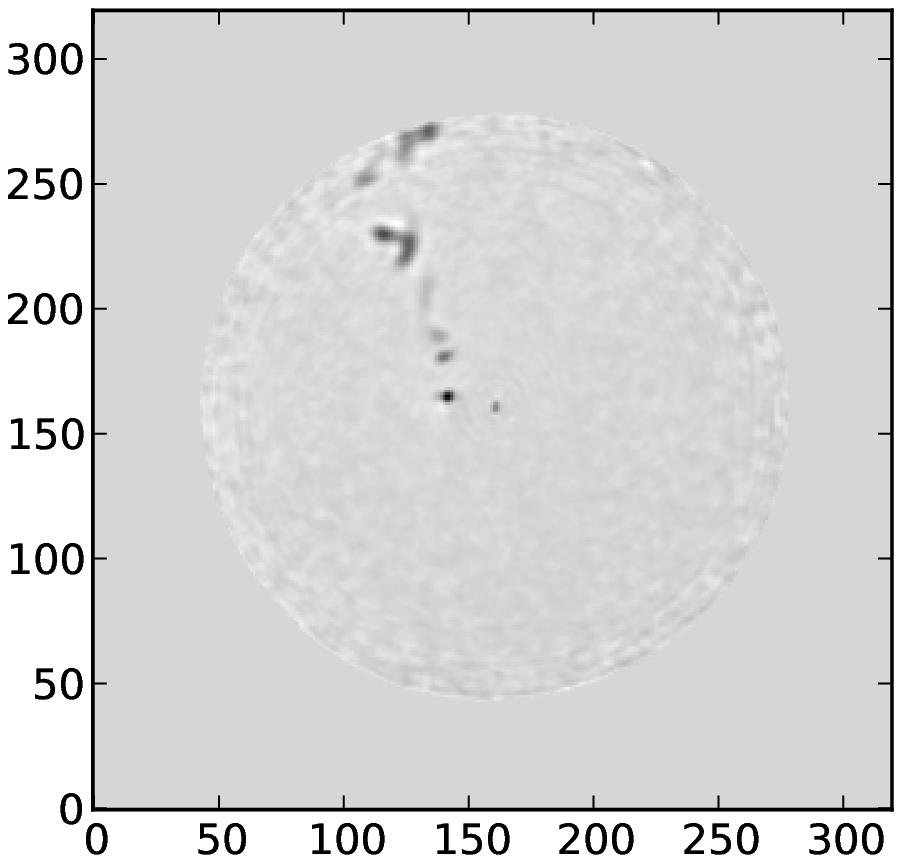}&
	\hspace{-7mm}\includegraphics[width=0.22\textwidth]{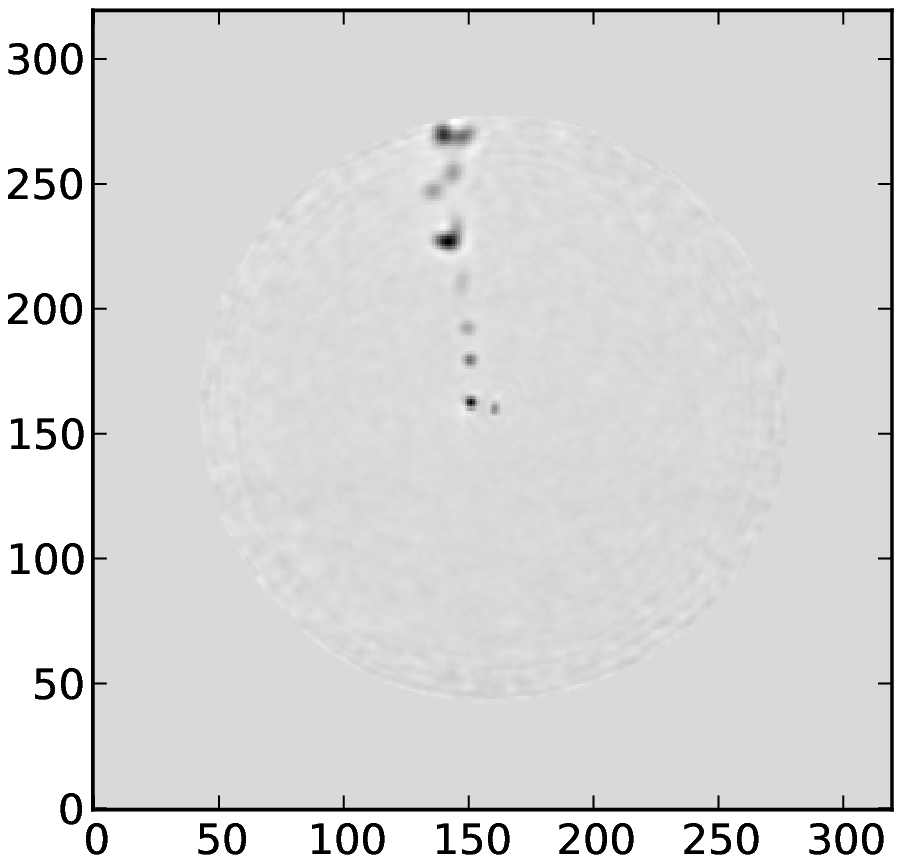}&
	\hspace{-7mm}\includegraphics[width=0.22\textwidth]{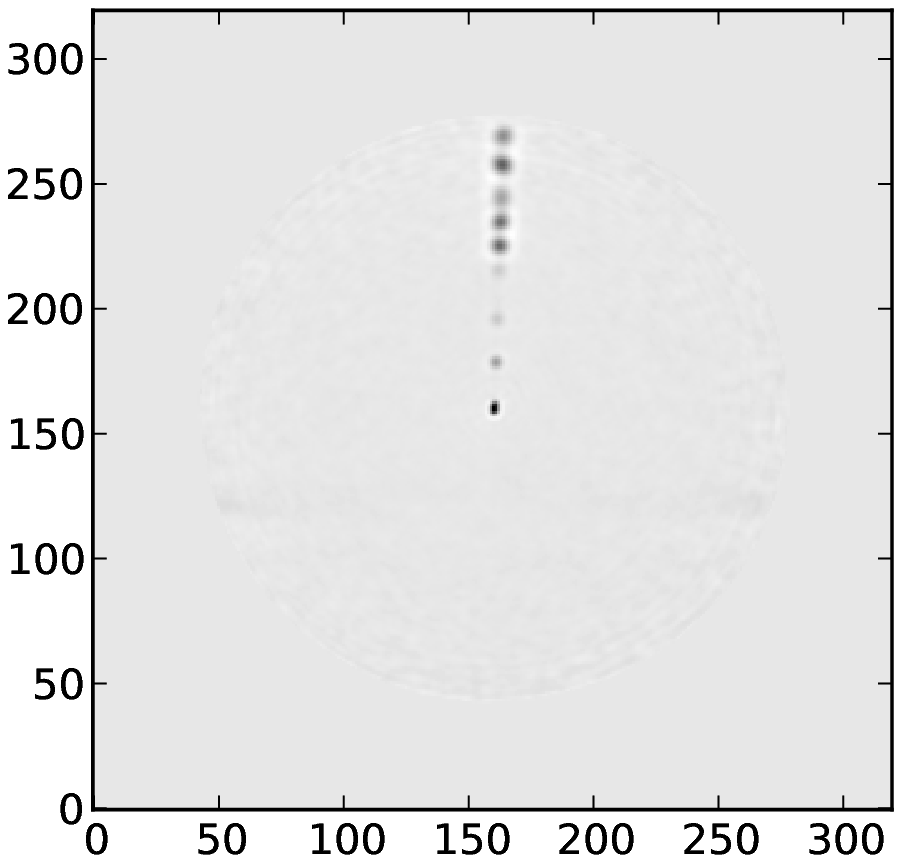}&
	\hspace{-7mm}\includegraphics[width=0.22\textwidth]{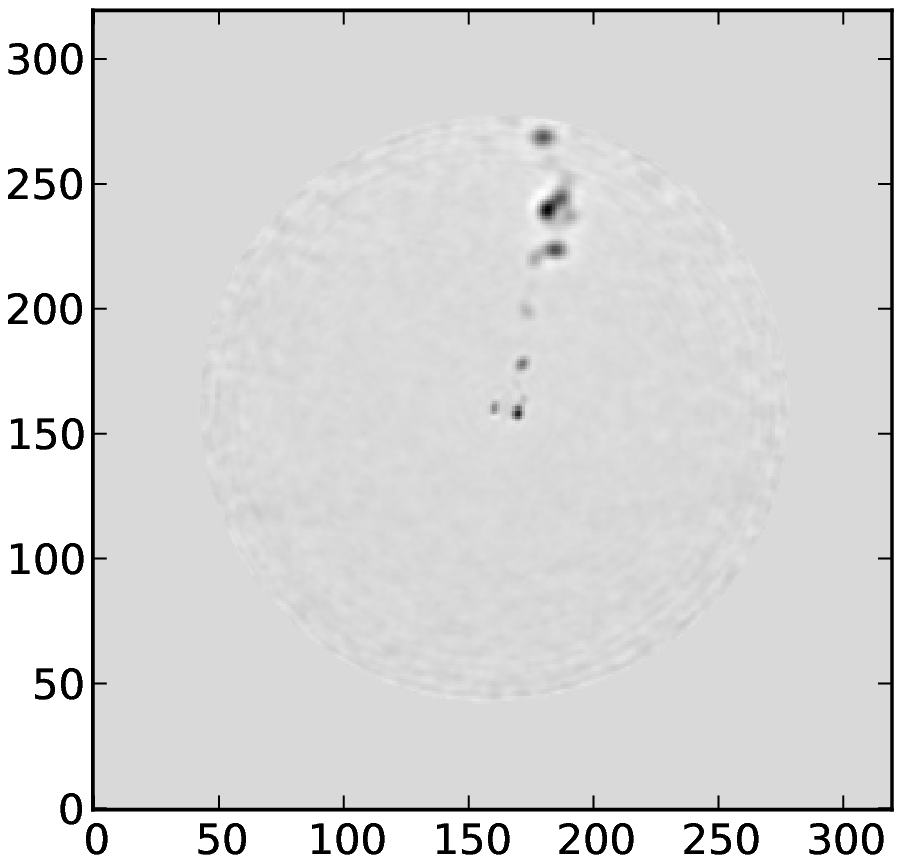}&
	\hspace{-7mm}\includegraphics[width=0.22\textwidth]{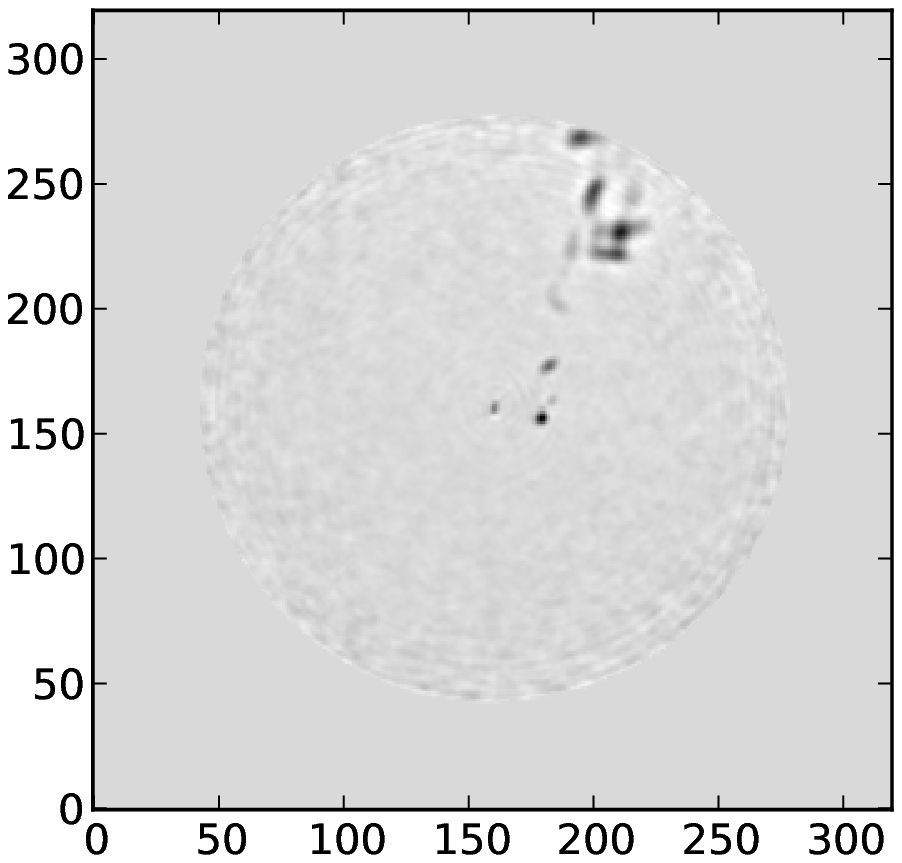}\\
	\hspace{-4mm}\mbox{\bf{(a)}}  & 
    \hspace{-7mm}\mbox{\bf{(b)}} &
    \hspace{-7mm}\mbox{\bf{(c)}}  & 
    \hspace{-7mm}\mbox{\bf{(d)}} & 
    \hspace{-7mm}\mbox{\bf{(e)}}  	
\end{array}$
\caption{Spatio--temporal cross--covariance functions of example on-sky data taken at a conjugate altitude of -2~km (inverted for clarity). The plots show cross--covariance functions generated with temporal delays equal to 1 frame ($\sim$10~ms) from -2 frames (a) to +2 frames (e). The case of no temporal delay is shown in (c). By examining the position of these peaks in subsequent frames the wind velocity (magnitude and direction) can be calculated.}
	\protect\label{fig:onskywind}
\end{figure*}

\begin{figure}
	\centering
	\includegraphics[width=0.25\textwidth]{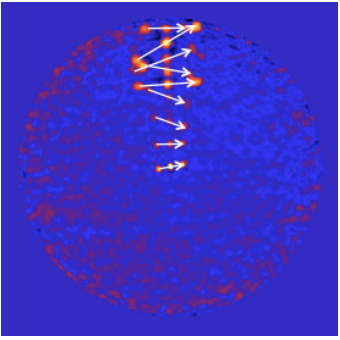}
	\caption{The sum of three consecutive spatio-temporal cross--covariance frames. We show the sum on one image to demonstrate the wind velocity estimation process. The arrows indicate the detected layers and velocities.}
	\protect\label{fig:xcov_sum3}
\end{figure}

In addition to fitting the spatial cross--covariance to recover the optical turbulence profile, if we assume `frozen flow' of the turbulence then wind velocity information can also be gleaned by examining the spatio-temporal cross--covariance. To do this we calculate the cross--covariance function for the two pupil images with increasing temporal offsets. By viewing the cross--covariance with increasing temporal delay we see that the covariance peaks from the turbulence will traverse the covariance function with a velocity corresponding to the velocity of the turbulence. Figure~\ref{fig:onskywind} shows the spatio-temporal cross--covariance functions for delays in the range -2 frames up to +2 frames. In this way the velocity of the turbulent layers can be estimated optically by tracking the covariance peaks through the spatio-temporal covariance function.
If we add the central three frames together (b, c and d from figure~\ref{fig:onskywind}) then it becomes easier to see the velocity of each layer (figure~\ref{fig:xcov_sum3}). 

\comment{, reducing the probability of false positive detections.}

To build a wind profile we assume frozen flow and implement a geometric algorithm. We make a least squares fit between equi-spaced covariance peaks in adjacent frames. We then do the same for several sets of three frames (positive and negative temporal offsets) so that we can detect layers even if they leave the scope of the cross--covariance function. To detect the velocity of a layer we require it to be seen in at least two sets of three frames, see \cite{Shepherd13} for more details. 

The wind velocity can be identified in this way by all SCIDAR systems \citep{Prieur04,Garcia_lorenzo06}, however the problem is simplified for Stereo-SCIDAR. All of the methods involve tracking the motion of the covariance peaks through the 2D covariance function, but with conventional SCIDAR systems there are three peaks for each turbulent layer making the extraction of velocity vectors complicated to automate. 

For the Stereo-SCIDAR on the INT, the estimated wind speed precision is $\pm 2.5\mathrm{ms}^{-1}$, corresponding to one pixel movement of the covariance peak in one frame. {It is possible to estimate the position of the coherence peak to sub-pixel precision. This would increase the precision of the wind velocity estimates, but is currently not implemented.}

The wind direction precision is wind speed dependent. For slow layers the angle is harder to determine due to the small number of pixels that the covariance peak will traverse. For the slowest layers the wind direction resolution is 10~degrees, for a typical wind speed of 10~m/s the wind direction resolution is 2.5~degrees. Here, we use 5 degrees (corresponding to 5~m/s wind speed) as a constant measurement precision of the wind direction for our analysis.

\begin{figure*}
\centering
    \includegraphics[width=0.95\textwidth,trim={1cm 1cm 1cm 0}]{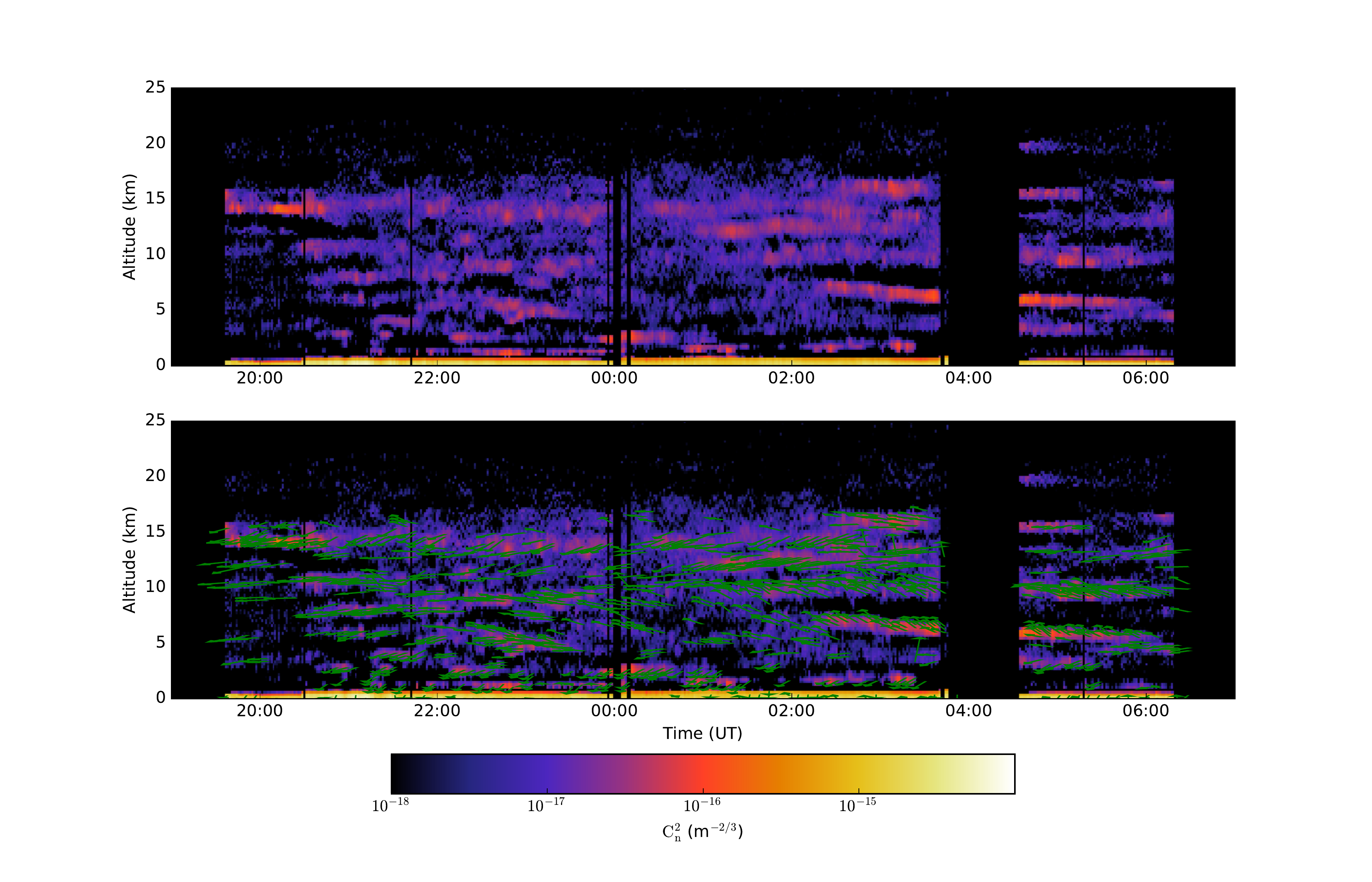}
	\caption{Example Stereo-SCIDAR profile sequence from the night of 9th October 2014 (top) and with wind vectors overlaid (bottom). Only a sub-sample of wind vectors are shown for clarity.}
	\protect\label{fig:nightprof}
\end{figure*}

Figure~\ref{fig:nightprof} shows an example recovered optical turbulence strength and velocity profile.

\changed{For the turbulence velocity identification we use the frozen flow assumption, i.e. that the turbulence does not evolve during the time it takes to cross the pupil. To calculate the velocity we use temporal delays of -5~ms to +5~ms. It has been shown that the frozen flow hypothesis is accurate over these short time scales (for example \citealp{Schoeck99,Guesalaga2014}).}
\comment{Gusting can blur covariance peak. Previous studies for slopes not scintillation}

\subsubsection{High Vertical Resolution Mode}

The altitude resolution (in terms of the minimum separation required to resolve two layers) of SCIDAR is altitude dependent and is given in equation~\ref{eq:dh}. For higher altitude layers this can be larger than 1~km. 

\changed{
In HVR mode \citep{Egner2007} the covariance peaks are tracked in the spatio-temporal covariance function. Using the frame rate and pixel size the turbulence velocity can be estimated as described above. The altitude of the turbulence can also be estimated by calculating the intersect of the peak's trajectory with the altitude axis in the spatial cross--covariance function (i.e. with no temporal delay). In this way the altitude of each of the layers can be estimated to a higher precision then the native profiling.
}
Using the HVR technique, the altitude resolution is no longer altitude dependent as layer altitudes can be ascertained to less than a Fresnel zone size. The HVR altitude resolution is given by \citep{Egner2007},
\begin{equation}
\mathrm{d}h = \frac{\cos{(\gamma)}D}{\theta n_\mathrm{pix}},
\label{eq:dh2}
\end{equation}
where $D$ is the telescope Diameter, $\theta$ is the angular separation of the \changed{double} stars and $n_\mathrm{npix}$ is the number of pixels across the pupil image. In typical conditions the altitude resolution of the La Palma system is $\sim$200~m. 

\changed{However, for Stereo-SCIDAR this has been further improved by fitting the trajectory of the covariance peak in the spatio-temporal covariance function to sub-pixel accuracy, enabling a sub-pixel estimate of the altitude. We record the altitude of the peak to 1/10$^{\mathrm{th}}$ of a pixel in the covariance function, approximately 20~m. To do this we fit all of the detected covariance peaks to a straight line and record the altitude given by this fit. An RMSE for the fit also allows us to judge the goodness of fit. This altitude resolution is sufficient for ELT-scale instrument performance modelling. 
}

\changed{It should be noted that a limitation of the HVR technique is that it can only separate layers if their altitudes and velocities are significantly distinct. It can not identify two layers moving with the same velocity within the altitude range of the Fresnel zone size in the spatio-temporal cross-covariance function, i.e. the covariance peaks must separate in the spatio-temporal cross-covariance function.
}

\subsubsection{Stereo-SCIDAR Data}
Table~\ref{tab:data} shows the volume of data that has been collected by Stereo-SCIDAR on the INT. In total 249 hours of profiles have been recorded over 28 nights. The results presented here are not meant to be representative of the La Palma site in general due to the limited data, clustered over a few months in two summers. The intention of this paper is to validate the wind velocity measurements by cross-comparisons of the three data sources. 
\begin{table}
\caption{Stereo-SCIDAR data volume}
\label{tab:data}
\begin{tabular}{@{}clccc}
\hline
Year & Month & Days & Hours of Data & Number of Profiles\\
\hline
2014 & March 	& 13 - 17 	& 32.9 & 533\\
	 & July		& 11 - 16 	& 49.7 & 1421\\
     & October	& 6 - 12 	& 61.9 & 1966\\
2015 & June		& 25 - 30 	& 47.9 & 1854\\
	 & July		& 1 		& 8.5 & 310 \\
	 & September& 29 - 30 	& 19.4 & 541\\
	 & October	& 1 - 5 	& 29.0 & 914\\
\hline
Totals:	&		& 28		& 249.3 & 7539\\
\hline
\end{tabular}
\end{table}

\changed{\subsubsection{Stereo-SCIDAR example cross-covariance functions}
\label{sect:xcov_examples}
In this section we show examples of the Stereo-SCIDAR spatio-temporal cross--covariance functions. Figure~\ref{fig:HVR_xcov} shows an example of the HVR technique. In this example we clearly see two turbulent layers at a very similar altitude with a small angular difference between them. The Stereo-SCIDAR data pipeline identifies two layers separated by 50~m. In this case it is likely that a zone of turbulent flow has occurred between two bodies of air. The temperature gradient at the top and the bottom of the turbulent zone has generated the changes in the refractive index and manifests as two turbulent layers. This supports earlier work by, for example, \cite{Coulman95}.
\begin{figure*}
\centering
$\begin{array}{c}
	\includegraphics[width=\textwidth]{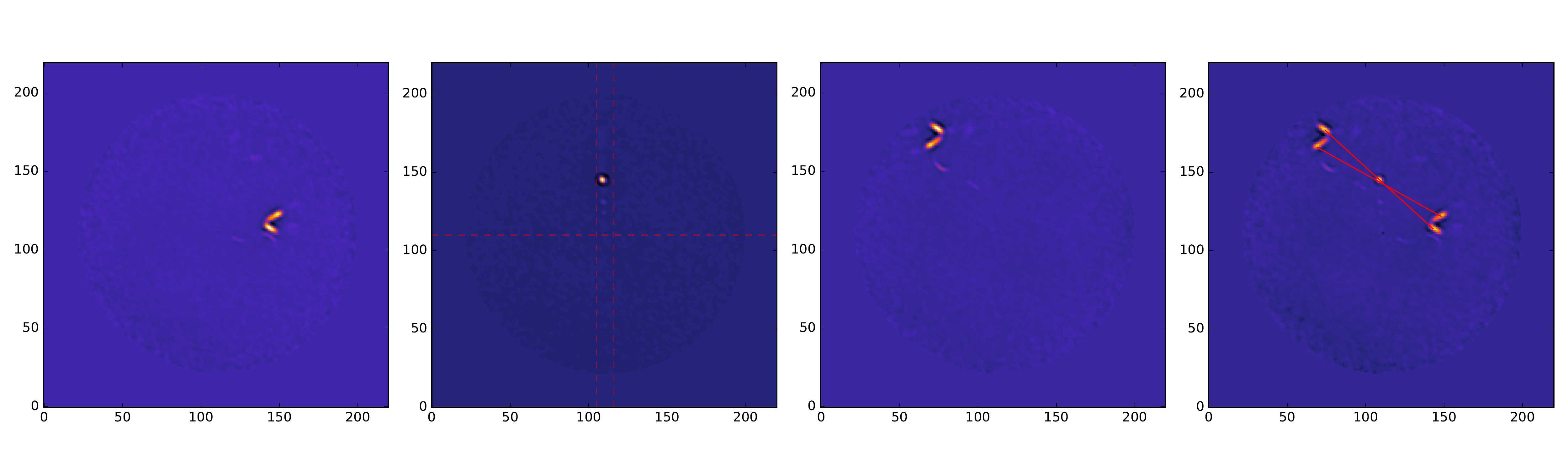}\\
	\begin{array}{cccc}
	\mbox{(a)} \hspace{3.7cm}& \mbox{(b)} \hspace{3.7cm}& \mbox{(c)} \hspace{3.7cm}& \mbox{(d)}
	\end{array}
\end{array}$
\caption{Example cross covariance function from 16th March 2014. The first three frames (a, b and c) are calculated with temporal offsets of -0.03 s, 0 s and +0.03 s respectively. (d) is the sum of the three frames with the turbulence velocity vector overlaid. The colour scale is different for each plot to enhance visibility. In this example there are clearly two layers of turbulence within a very small altitude range moving with a small angular differential (approximately 10 degrees). The HVR technique can separate these layers. In this case the automated script identified two layers at separated by 50 metres in altitude with wind speeds of 25 and 30 m/s and directions of 247 and 259 degrees.}
\label{fig:HVR_xcov}
\end{figure*}

Figure~\ref{fig:Arc_xcov} shows another typical example. In this case an arc can be seen in the spatio-temporal covariance function. This suggests a zone of turbulence with a dispersion of wind velocities and can be explained by a zone of turbulent flow between two bodies of air which are moving with a difference in their direction. In this case the motion of the turbulence is distributed between the two. The HVR Stereo-SCIDAR pipeline is not designed to identify these arcs and so fits multiple independent layers to the arc. In this case it identifies 8 turbulence vectors within a 200~m altitude range.
\begin{figure*}
\centering
$\begin{array}{c}
	\includegraphics[width=1.\textwidth]{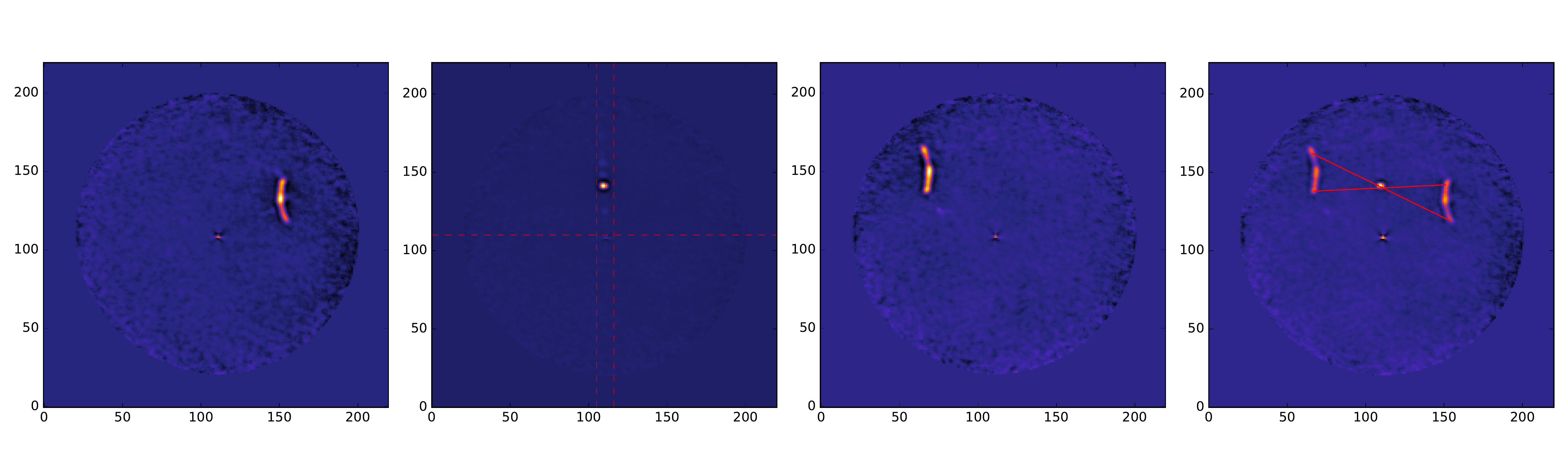}\\
	\begin{array}{cccc}
	\mbox{(a)} \hspace{3.7cm}& \mbox{(b)} \hspace{3.7cm}& \mbox{(c)} \hspace{3.7cm}& \mbox{(d)}
	\end{array}
\end{array}$
\caption{Example cross covariance function from 17th March 2014. The first three frames (a, b and c) are calculated with temporal offsets of -0.03 s, 0 s and +0.03 s respectively. (d) is the sum of the three frames with the range of the possible turbulence velocity vectors overlaid. The colour scale is different for each plot to enhance visibility. These arcs are often seen in the spatio-temporal cross covariance function and show that a range of velocities can be seen within a single turbulent layer. In this case the arc covers a range of angles of approximately 30 degrees. The Stereo-SICDAR data pipeline identified 8 independent vectors corresponding to layers within 200~m (3050 m to 3350 m) with a spread of 10 m/s (20 m/s to 30 m/s) and 30 degrees (270 degrees to 300 degrees).}
\label{fig:Arc_xcov}
\end{figure*}

\comment{
XXX
Weak turbulence??
XXX}

The examples in figures~\ref{fig:HVR_xcov} and \ref{fig:Arc_xcov} were chosen as clear examples to show the effects that we see regularly. In figure~\ref{fig:xcov_example} we show an example of what the cross--covariance functions usually look like. It can be seen that the turbulent structure of the atmosphere is complex. We see a combination of double layers and arcs at all altitudes.
\begin{figure*}
\centering
$\begin{array}{c}
	\includegraphics[width=1.\textwidth]{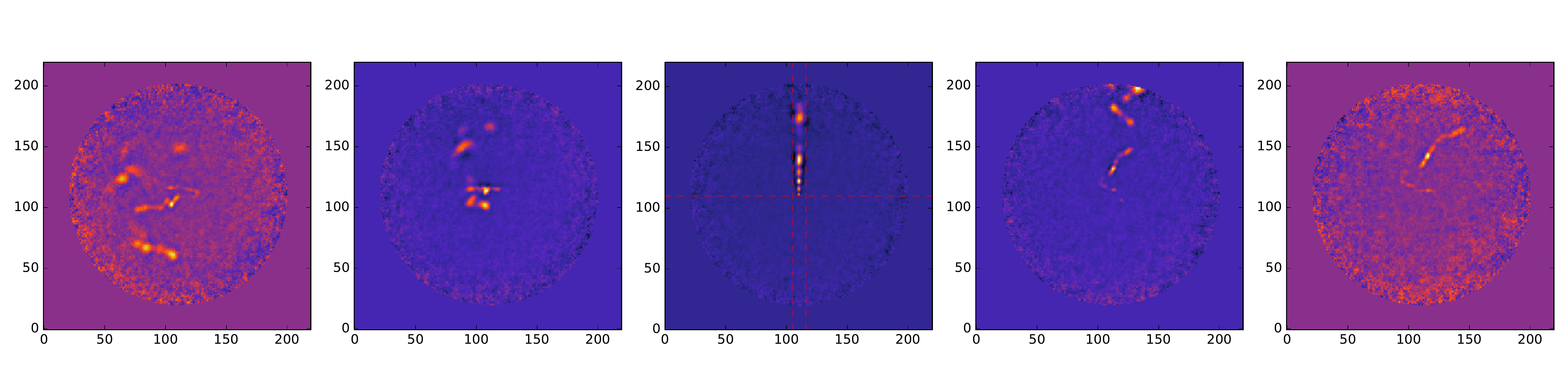}\\
	\begin{array}{ccccc}
	\mbox{(a)} \hspace{2.8cm}& \mbox{(b)} \hspace{2.8cm}& \mbox{(c)} \hspace{2.8cm}& \mbox{(d)} \hspace{2.8cm}& \mbox{(e)}
	\end{array}
\end{array}$
\caption{Typical example cross covariance function from 17th March 2014 showing a complex turbulence structure with arcs and double layers occurring at all altitudes. The cross-covarinace functions are calculated with temporal offsets from -0.04 to +0.04s. The central frame (c) shows the case with no temporal offset. The colour scale is different for each plot to enhance visibility.}
\label{fig:xcov_example}
\end{figure*}
}

\subsection{Radiosonde}
Balloon borne radiosondes are released twice daily from Valle de Guimar, Tenerife. The launches are at 12:00 and 00:00 UT, here we only use the data from the 00:00 UT launch, {i.e. one launch per night}. Valle de Guimar, Tenerife, is approximately 150~km from the INT on La Palma. The launch site is also at sea level, 2330~m below the INT. 

The average ascension rate of the radiosonde is 300~m per minute \citep{NOAA1997}. This means that it will take approximately 85~minutes to probe the atmosphere up to 25~km. In this time, assuming a high average horizontal wind speed of 30~ms$^{-1}$ the balloon could drift up to 150~km from the launch site. Therefore, the radiosonde is not probing the same line of sight as the telescope and the GFS forecast. We do not expect the correlation to be perfect, however for higher altitudes, away from any local surface effects, the meteorological conditions should be similar. 

The quoted measurement precision of the wind speed and direction is 3~m/s and 5~degrees respectively \citep{NOAA1997}. {The altitude resolution of the radiosondes is variable but tends to be a few hundred metres.}

\subsection{GFS}
The GFS model from the National Oceanic and Atmospheric Administration \citep{NOAA2015} provides global forecasts of meteorological parameters, including the vertical profile of the wind speed and direction. The GFS model has a horizontal resolution of 0.5~degree (55.4~km N-S, 48.2~km E-W at the latitude of La Palma) for 2014 data and 0.25~degree (27.7~km N-S, 24.1~km E-W at the latitude of La Palma) for 2015 onwards. The vertical resolution has 25~mb resolution from 1000 to 900~mb and then 50~mb resolution down to 50~mb. The model is {updated} every six~hours and provides forecasts for the next 16 days in three hour time intervals.

We compare two versions of the GFS forecast, the first is the forecast produced at 18:00UT for 00:00UT (GFS +6), the same time as the radiosonde launch. The second forecast is produced at 00:00UT for the current time (GFS +0). This short time-scale forecast should be the most accurate approach, but we include the 6 hour forecast version for interest.
 
General circulation models (GCM), such as GFS, have been {previously used to provide free atmosphere (above the ground layer)} wind velocity profiles (for example, \citealp{Hagelin2010}, and references therein). It is known that these models can be unreliable in the lower atmosphere, where local geography can influence the climatic parameters. The meso-scale approach of \cite{Masciadri2013a}, is able to provide these parameters, including the ground layer, with 2~minute temporal and 100~m spatial resolution.

Here, we only use the wind velocity information as it is presented in the GFS datasets, no further data manipulation is performed. Therefore, unlike more sophisticated models (the MOSE project for example, \citealp{Masciadri2013a}), we limit ourselves to the temporal and spatial resolution of the model provided. For this reason, it should also be noted that we are not proposing to use the GFS wind velocities directly in the AO control optimisation, or for any other application, but only to validate the Stereo-SCIDAR measurements and to show that the wind velocity can be equated to the optical turbulence velocity.

\section{Comparisons}
\label{sect:comp}
For these comparisons, only Stereo-SCIDAR profiles recorded between the hours of 23:30 and 00:30 UT are included to ensure that the data that we compare is within half an hour of the radiosonde launch and GFS forecast time. Due to the flight time of the radiosonde a better correlation might be found with Stereo-SCIDAR data from 00:00 to 01:00 UT, however the GFS forecasts are published for 00:00 UT and so we have chosen to use midnight as the centre of all comparisons. \changed{Due to the different altitude resolutions of the three systems we only compare measurements that are within 100~m in altitude of each other.} The data shown here is for all altitudes above the observatory level, including the ground. If the data from the first kilometre above the observatory is ignored no significant difference is seen in any of the comparisons.

\begin{figure*}
\centering
$\begin{array}{cc}
	\includegraphics[width=0.5\textwidth,trim={1.5cm 0 1cm 0}]{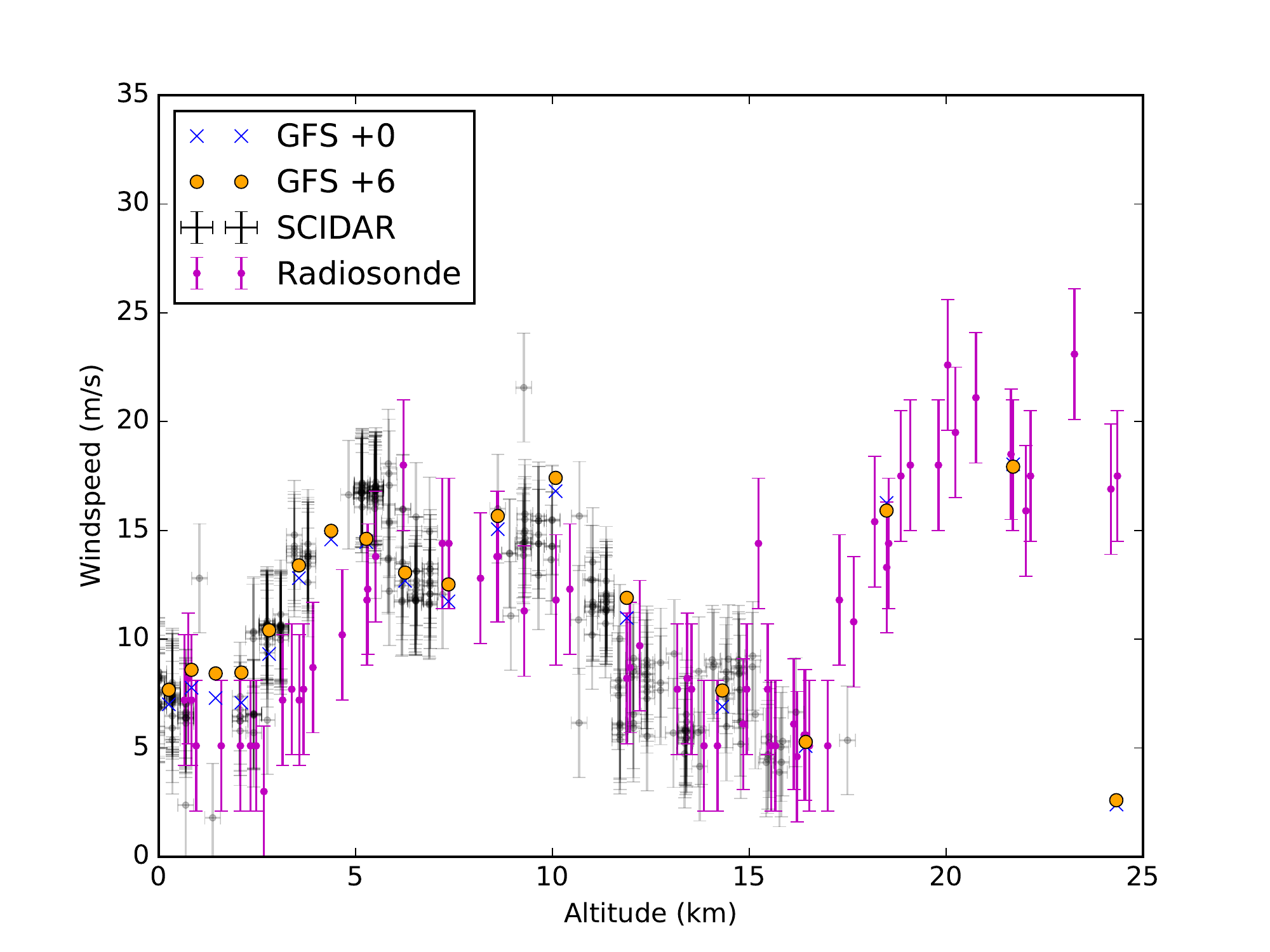}&
    \includegraphics[width=0.5\textwidth,trim={1.5cm 0 1cm 0}]{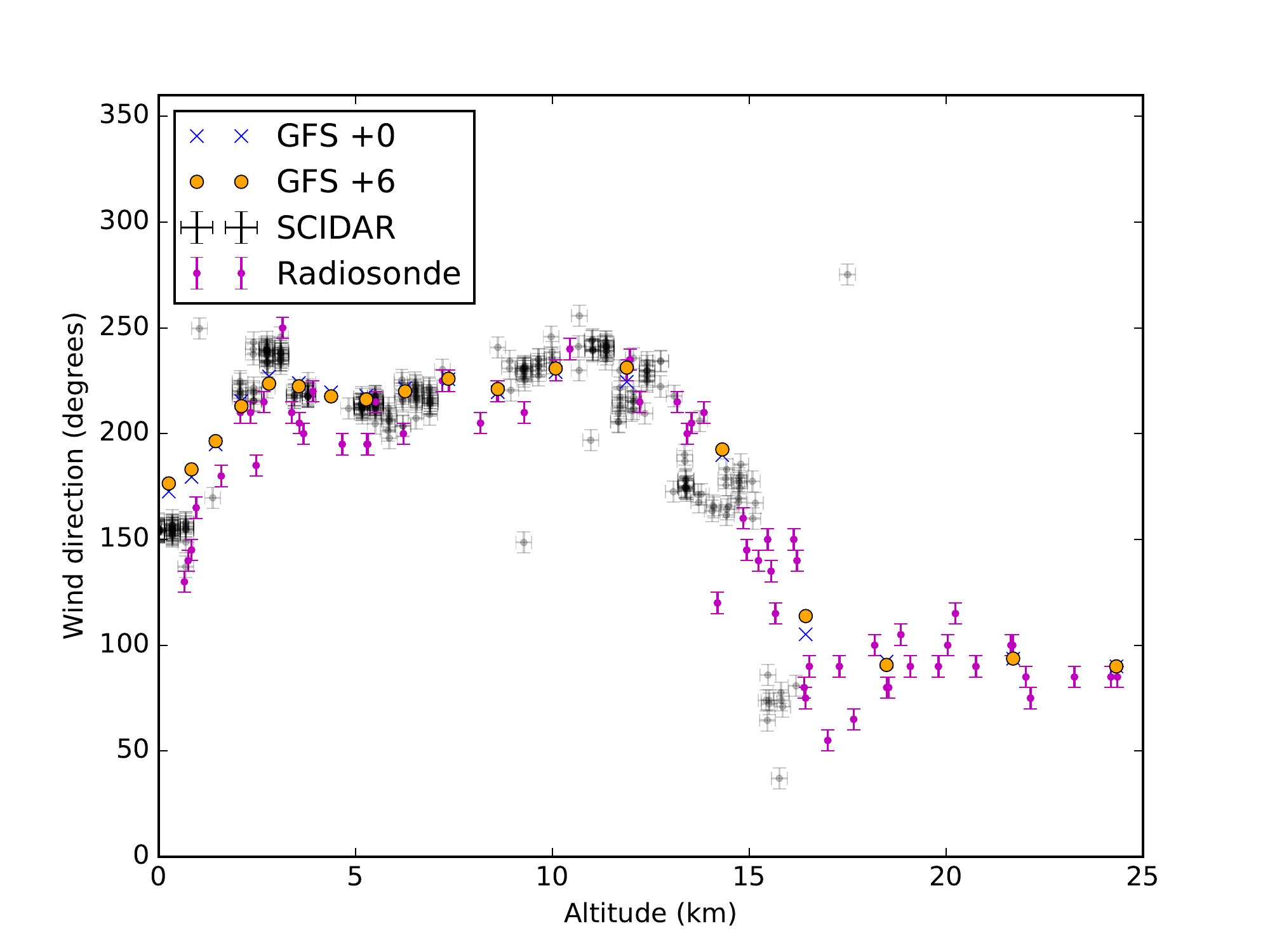}
\end{array}$
\caption{An example velocity profile for wind speed (left) and direction (right) from the Stereo-SCIDAR on the INT, GFS model and radiosonde measurements for 00:00 UT the night of 2014/07/15.}
\label{fig:wind_example}
\end{figure*}

An example velocity profile for 2014/07/15 is shown in figure~\ref{fig:wind_example}. On this particular night we have data from the GFS forecast, the radiosonde and the Stereo-SCIDAR.

\begin{figure*}
\centering
$\begin{array}{cc}
	\includegraphics[width=0.5\textwidth,trim={1.5cm 0 1cm 0}]{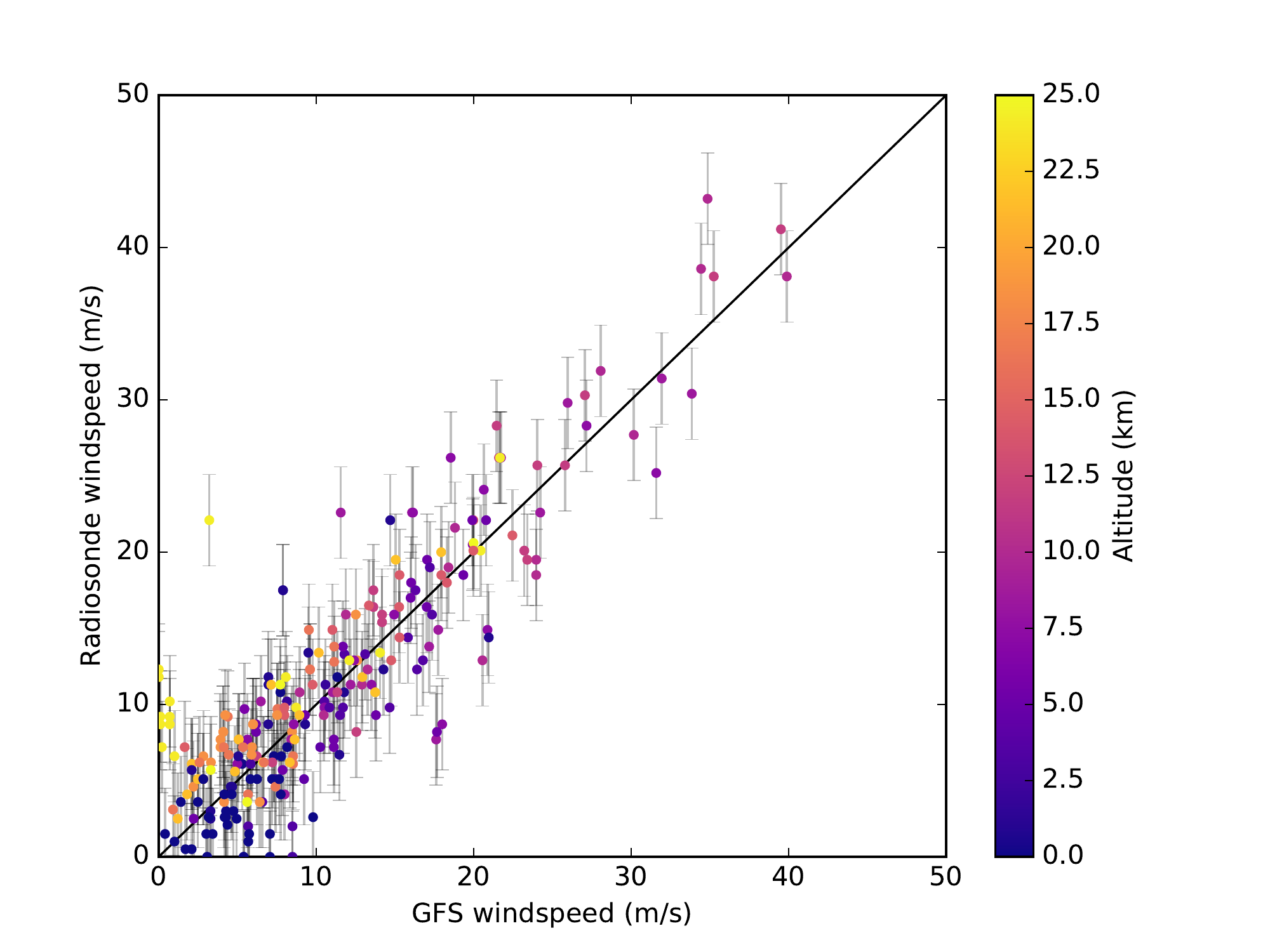}&
    \includegraphics[width=0.5\textwidth,trim={1.5cm 0 1cm 0}]{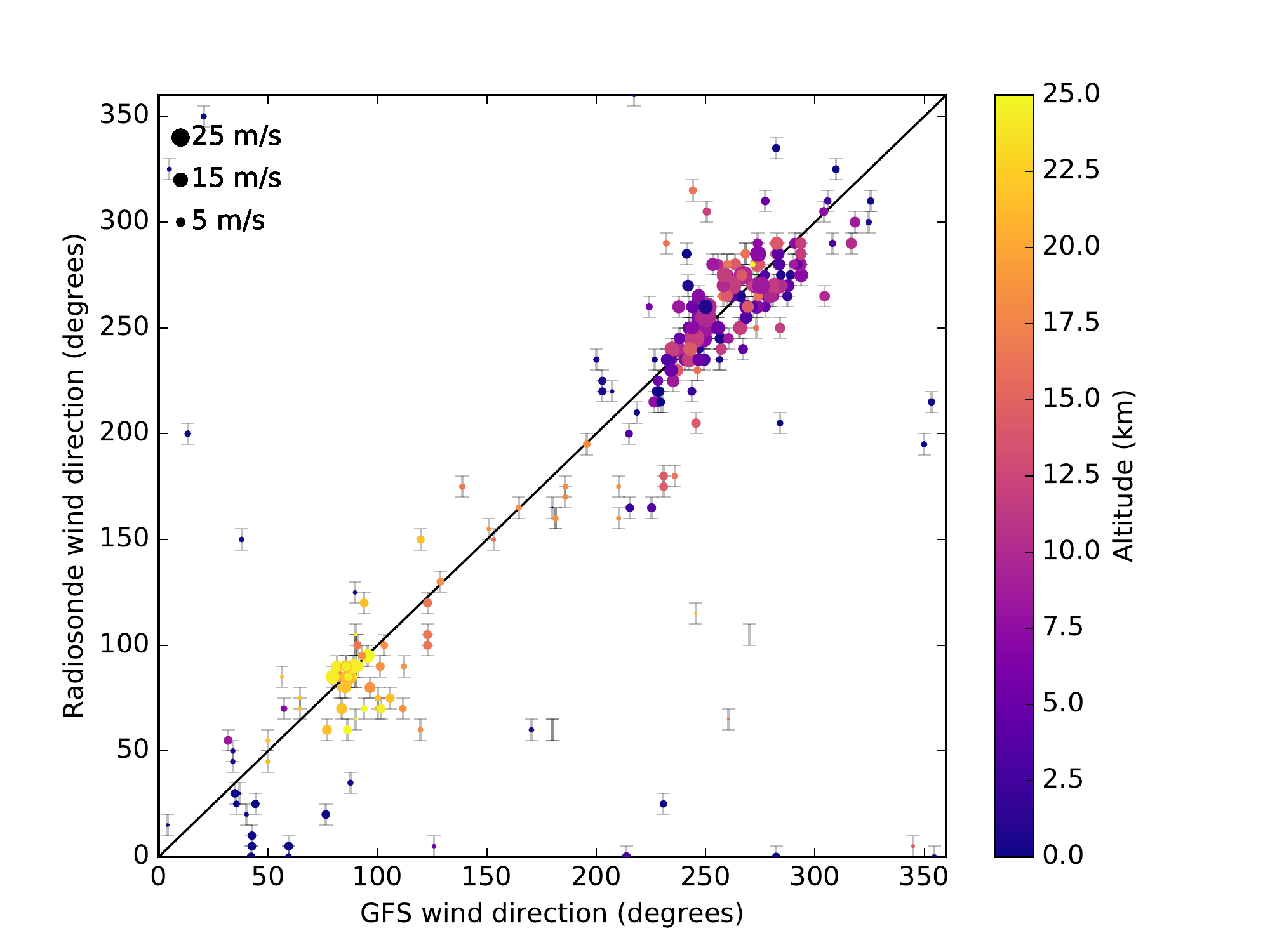}
\end{array}$
\caption{Comparison of wind speed (left) and wind direction (right) from radiosonde and GFS. The colour indicates the altitude of the turbulence. For the comparison of the direction (right) the size of the point indicates the wind speed.}
\label{fig:RSGFS}
\end{figure*}

\begin{figure*}
\centering
$\begin{array}{cc}
	\includegraphics[width=0.5\textwidth,trim={1.5cm 0 1cm 0}]{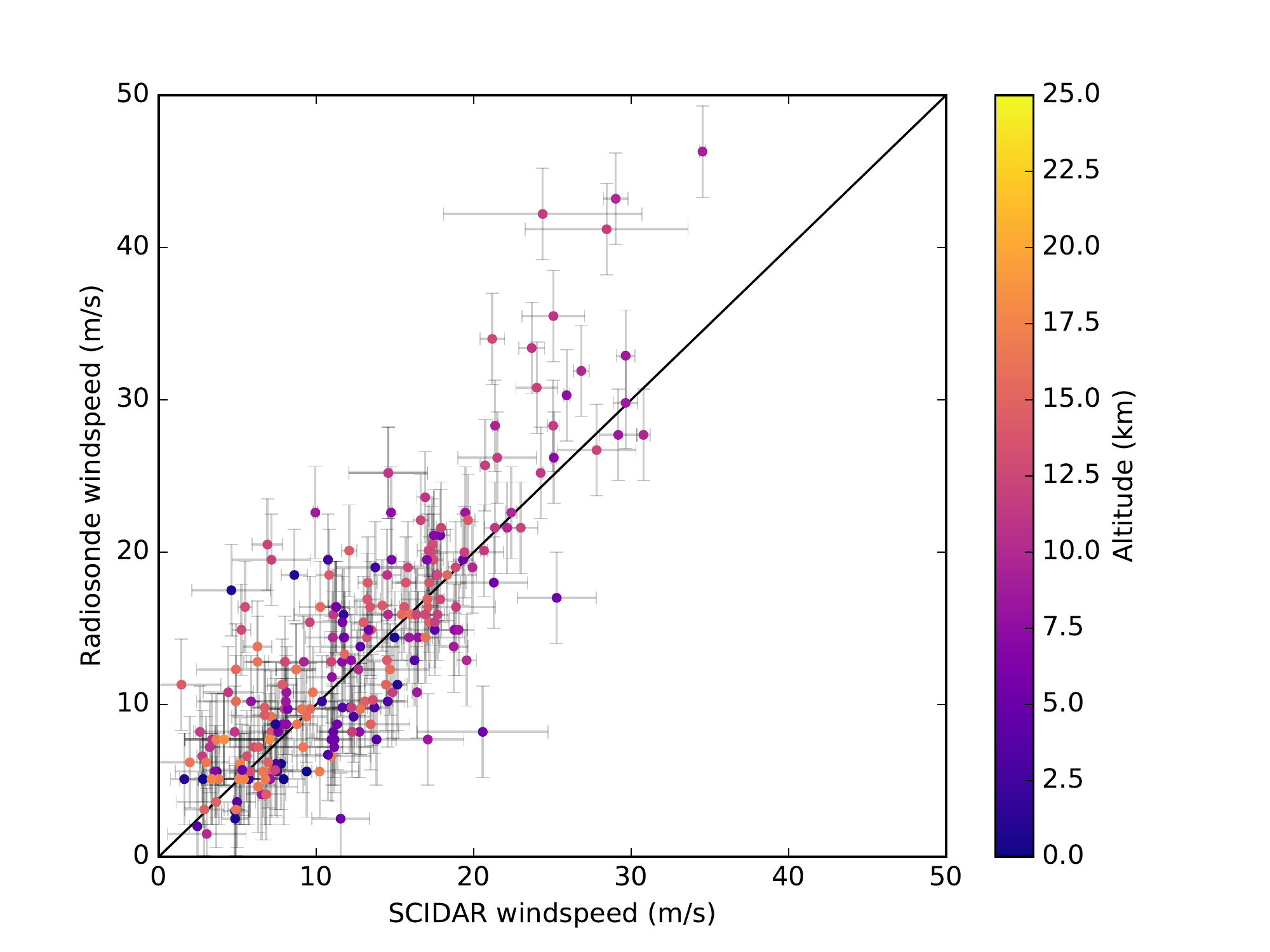}&
    \includegraphics[width=0.5\textwidth,trim={1.5cm 0 1cm 0}]{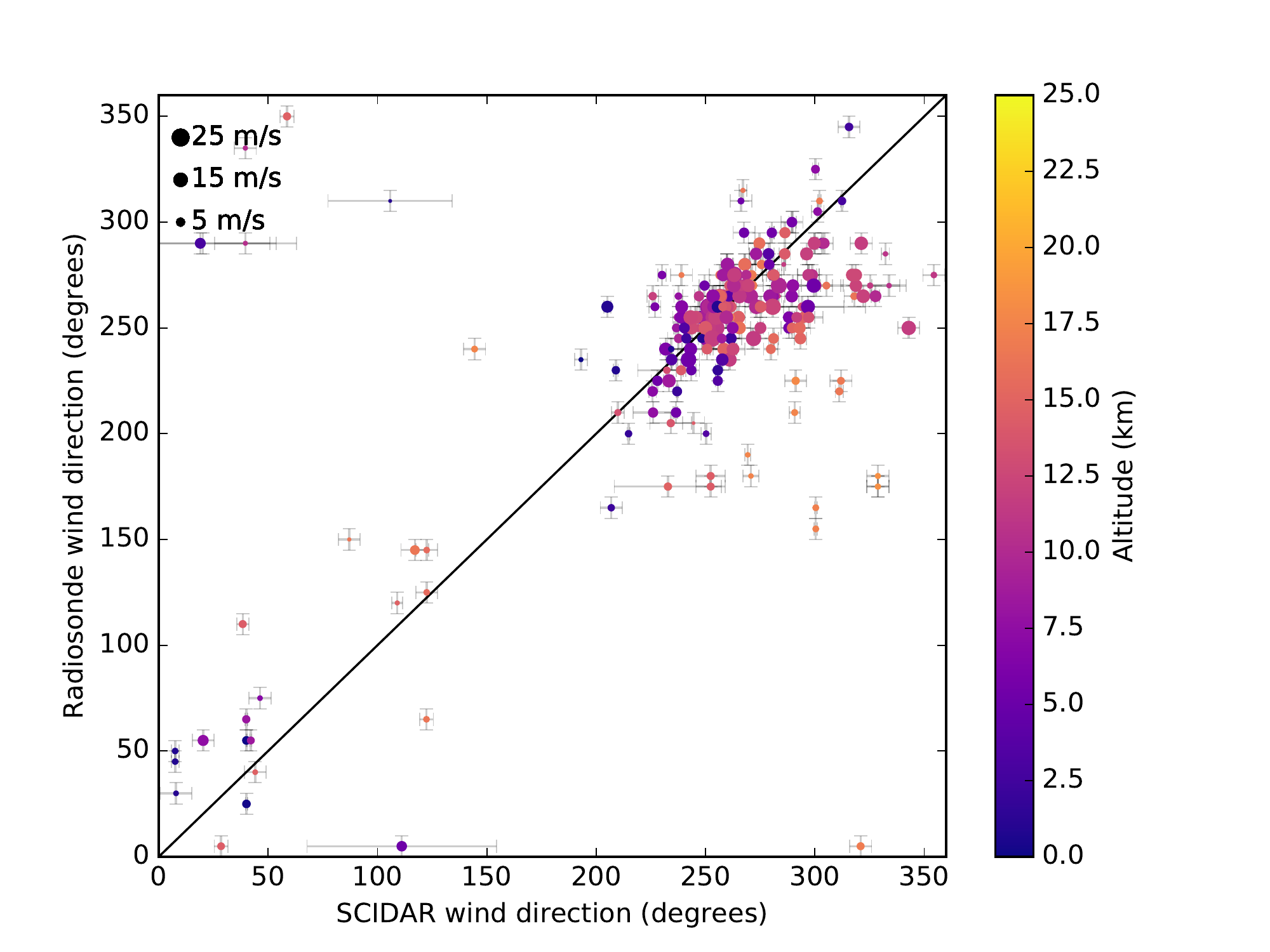}
\end{array}$
\caption{Comparison of wind speed (left) and wind direction (right) from radiosonde and Stereo-SCIDAR. The colour indicates the altitude of the turbulence. For the comparison of the direction (right) the size of the point indicates the wind speed.}
\label{fig:RSSCIDAR}
\end{figure*}

\begin{figure*}
\centering
$\begin{array}{cc}
	\includegraphics[width=0.5\textwidth,trim={1.5cm 0 1cm 0}]{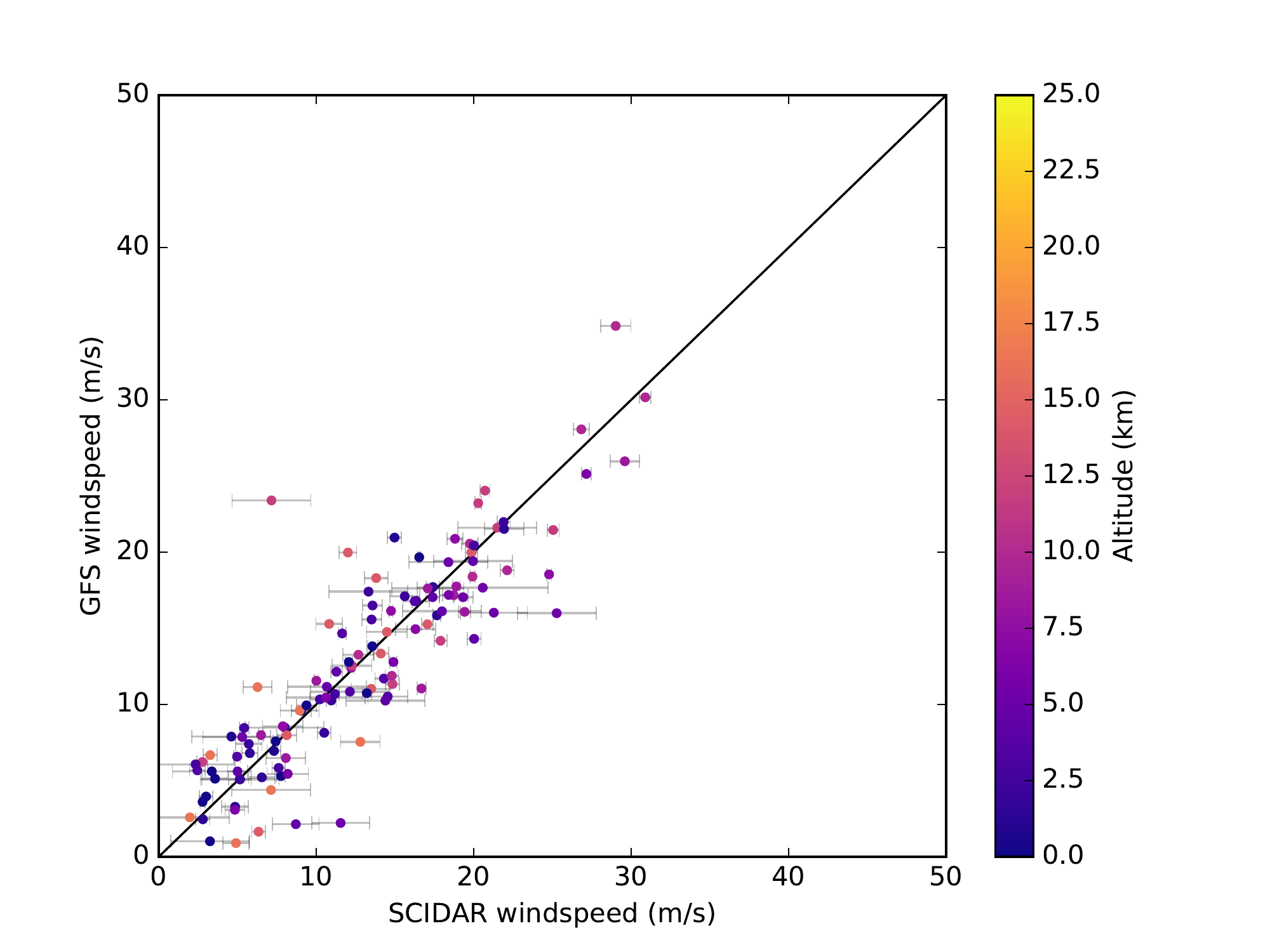}&
    \includegraphics[width=0.5\textwidth,trim={1.5cm 0 1cm 0}]{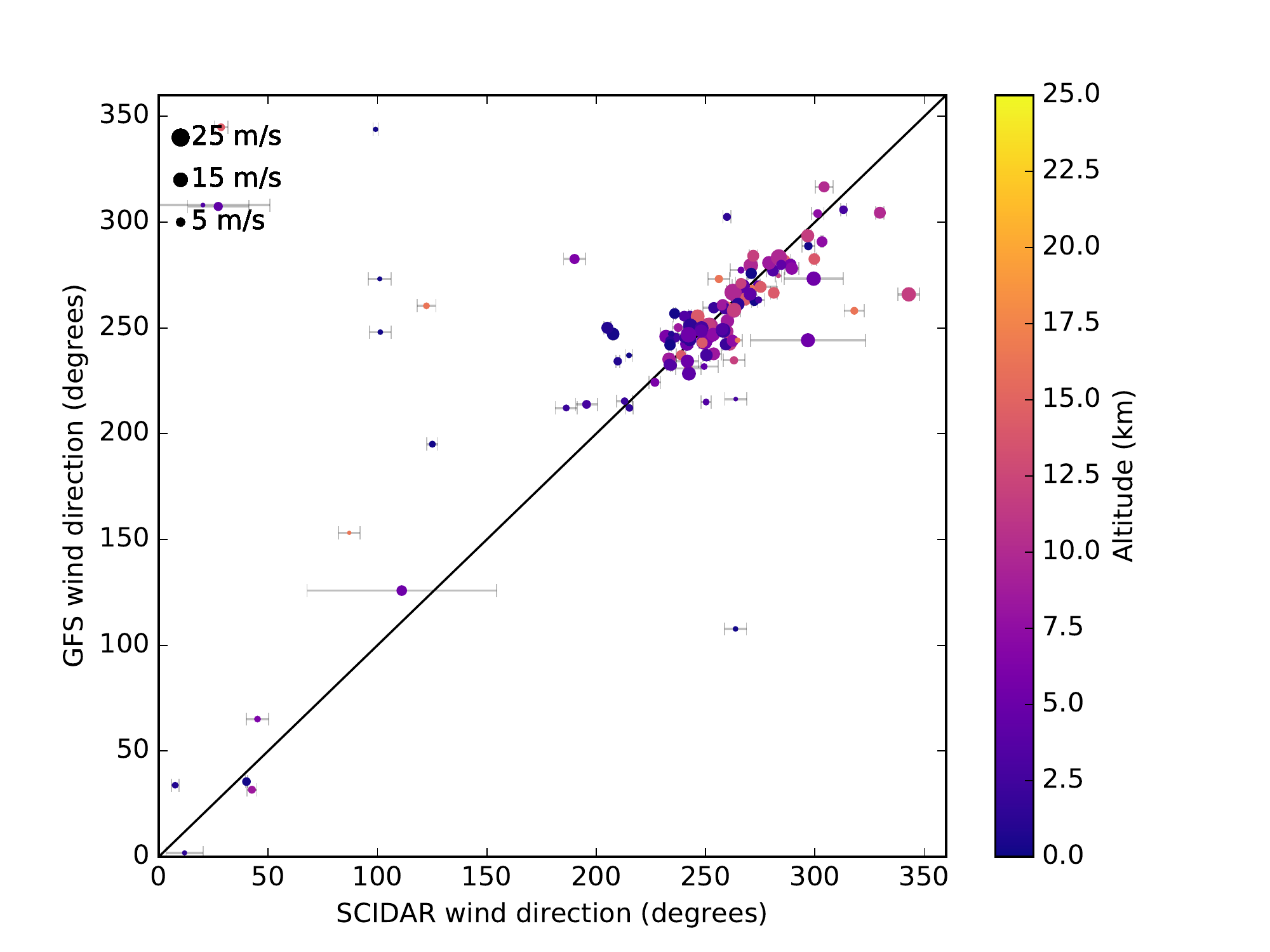}
\end{array}$
\caption{Comparison of wind speed (left) and wind direction (right) from GFS and Stereo-SCIDAR. The colour indicates the altitude of the turbulence. For the comparison of the direction (right) the size of the point indicates the wind speed.}
\label{fig:GFSSCIDAR}
\end{figure*}

Figures~\ref{fig:RSGFS}, \ref{fig:RSSCIDAR} and \ref{fig:GFSSCIDAR}
show the scatter plots for different combinations of the data sources. The colour of the marker indicates the altitude of the turbulent layer. \changed{We first compare the radiosonde measurements with the GFS forecast for our dataset to confirm that these sources agree and can be used to validate the SCIDAR velocities.}

\comment{In most cases there is no altitude dependence on the correlation between the various data sources, with the exception of the comparison of the wind direction from the radiosonde and GFS at observatory level. In figure~\ref{fig:RSGFS} we see that there is large scatter in the radiosonde wind direction at low altitudes as compared to the GFS.}

The large error bars for the SCIDAR data show a large variability in some of the wind velocity measurements. For the comparisons of the wind velocity, speed and direction, measurements from Stereo-SCIDAR are averaged over the one hour period. The error bars are then set to the standard error of the measurements. These error bars appear large in comparison to the quoted measurement resolution and therefore show a large variability in some of the wind velocity measurements. This could be due to actual wind variability over the comparison period and to turbulent gusting, a known phenomenon in turbulent flow \citep{Boettcher2003}. Also, note that the wind direction is wrapped in the plots and therefore points in opposing corners away from the main trend should wrap around at 360~degrees.

Table~\ref{tab:corr_table} shows the Pearson correlation values from each combination of data sources.\comment{ The $\chi^2$ value for the comparisons are also show in the plot. The relative errors are included in the $\chi^2$ analysis and so this is perhaps a better measure of correlation than the Pearson correlation. Both are included for completeness.} The correlation for all combinations are high, indicating a good \changed{agreement} between each data source.\comment{ This is particularly true for the wind speed, which shows a higher correlation than the wind direction for all combinations of datasets.}

The correlation for all of the datasets with GFS +0 is slightly higher than that of GFS +6, and the correlation between the GFS +0 and GFS +6 is extremely high. This indicates that the six hour forecast is generally very precise. {Only the GFS +0 data is shown in the figure~\ref{fig:GFSSCIDAR} for clarity.}

\comment{
\begin{table}
\caption{Correlation values for all combinations of data sources for wind speed, $W_{\mathrm{s}}$, (m/s) and direction, $W_\theta$, (degrees).}
\label{tab:corr_table}
\begin{tabular}{@{}llcc}
\hline
& & \multicolumn{2}{c}{Correlation}\\
Data source 1 & Data source 2 & $ W_{\mathrm{s}}$ & $W_\theta$\\
\hline
Radiosonde & GFS +0 & 0.86 & 0.93 \\
Radiosonde & GFS +6 & 0.84 & 0.92 \\
Radiosonde & SCIDAR & 0.75 & 0.84 \\
SCIDAR & GFS +0 & 0.85 & 0.91 \\
SCIDAR & GFS +6 & 0.86 & 0.89 \\
GFS +0 & GFS +6 & 0.99 & 0.97\\
\hline
\end{tabular}
\end{table}
}
\begin{table}
\caption{Correlation values for all combinations of data sources for wind speed, $W_{\mathrm{s}}$, (m/s) and direction, $W_\theta$, (degrees).}
\label{tab:corr_table}
\begin{tabular}{@{}llcc}
\hline
& & \multicolumn{2}{c}{Correlation}\\
Data source 1 & Data source 2 & $ W_{\mathrm{s}}$ & $W_\theta$\\
\hline
Radiosonde & GFS +0 & 0.90 & 0.96 \\
Radiosonde & GFS +6 & 0.88 & 0.96 \\
Radiosonde & SCIDAR & 0.82 & 0.82 \\
SCIDAR & GFS +0 & 0.90 & 0.93 \\
SCIDAR & GFS +6 & 0.88 & 0.86 \\
GFS +0 & GFS +6 & 0.99 & 0.97\\
\hline
\end{tabular}
\end{table}

The mean difference, or bias, and the root-mean-square-error (RMSE) in the wind velocity values are shown in table~\ref{tab:bias_table}. The bias and RMSE values between the radiosonde and GFS model are consistent with those stated by \cite{Masciadri2013b} for radiosonde and ECWMF (an alternative GCM). We see a very low bias for the wind speed $\sim \pm$0.5~m/s, for all comparisons and a bias of up to $\sim$5~degrees for the wind direction. 

\comment{The fractional bias and the RMSE of the wind direction is consistently higher than the wind speed. This is partly because wind direction identification of slow turbulent layers is inherently more difficult due to the fewer pixels covered by the covariance peak in the sample time. In addition, the correlation calculation does not take into account the greater measurement uncertainty of wind direction from both the Stereo-SCIDAR and the radiosonde. It is also possible that gusting in the turbulence is more variable in direction than in magnitude.}

\changed{
The RMSE for the turbulence speed is of the same order as the measurement precision. However, the RMSE for the turbulence direction is larger than the expected measurement error. This can be explained by the fact the the Stereo-SCIDAR measurements given are the median velocity over an hour of observations whereas the GFS and radiosonde estimates were instantaneous. This is demonstrated by the error bars in the figures. As the RMSE for wind speed is low, this would suggest that there is more variation in direction over the hour. Another explanation is that the SCIDAR measures the velocity of the optical turbulence whereas the radiosonde and GFS give estimates of the wind velocity in discrete altitude bins. Turbulent zones with velocity dispersion in a small altitude range, as shown in section~\ref{sect:xcov_examples} will result in a difference between the optical turbulence direction and the Radiosonde / GFS discrete wind direction resulting in scatter in the above figures.

\comment{The RMSE for the comparison of the radiosonde and GFS is also higher than the measurement precision. This is probably due to the time difference and spatial displacement between the two estimates.}

We also note that outlying data points for the wind direction comparisons tend to be from low wind speed (<5~m/s). This is because wind direction identification of slow turbulent layers is inherently more difficult due to the fewer pixels covered by the covariance peak in the sample time.
}

\begin{table}
\caption{Bias and RMSE values for all combinations of data sources for wind speed (m/s) and direction (degrees).}
\label{tab:bias_table}
\begin{tabular}{@{}llcccc}
\hline
& & \multicolumn{2}{c}{Bias}& \multicolumn{2}{c}{RMSE}\\
Data source 1 &Data source 2 & $ W_{\mathrm{s}}$& $W_\theta$ & $ W_{\mathrm{s}}$  & $W_\theta$ \\
& & (m/s) & (degrees) & (m/s) & (degrees)\\
\hline
Radiosonde & GFS +0 & -0.1 & 3.5 & 2.2 & 17.0 \\ 
Radiosonde & GFS +6 & 0.8 & 5.6 & 2.5 & 17.4 \\ 
SCIDAR & Radiosonde &  0.6 & -1.9 & 3.3 & 23.5 \\
SCIDAR & GFS +0 & -0.8 & -2.6 & 1.9 & 12.5 \\
SCIDAR & GFS +6 & -0.7 & -1.5 & 2.0 & 19.9 \\
\comment{GFS +0 & GFS +6 & 0.1  & 0.5  & 0.6 & 20.3\\}
\hline
\end{tabular}
\end{table}

\section{Conclusions}
\label{sect:conc}
We have compared the wind velocity (speed and direction) from the GFS CGM model and from balloon borne radiosonde with turbulence velocity measurements from Stereo-SCIDAR on the INT, La Palma. As all three data sources agree with a high degree of correlation and a low bias we can say that the wind velocity, as measured by Radiosonde and GFS, is consistent with the turbulence velocity as measured by Stereo-SCIDAR. 

The turbulence velocity dispersion measured by the Stereo-SCIDAR for some turbulent zones provides an insight into the structure within turbulent flow. The dispersion is not seen by the radiosonde or the numerical model, GFS. We often see significant velocity dispersion and double layers which manifest in the comparison as a large RMSE when compared to the measurement precision.

\comment{
This is a significant result as it confirms that atmospheric models such as GFS can indeed reproduce turbulence velocity and not just wind velocity. This is important as astronomical optical systems are sensitive to the velocity of the optical turbulence. XXX velocity dispersion XXX
}

This work validates the use of the Stereo-SCIDAR real-time automated turbulence velocity identification algorithm. This result is significant because it justifies the use of Stereo-SCIDAR in the high-vertical resolution mode, where differential velocity vectors can be used to increase the altitude resolution of the turbulence profiles. This is currently the only way of achieving the same altitude resolution as the future generation of ELTs on 1 to 2~m class telescopes. 

\comment{
In addition, the turbulence velocity identification is the only way of validating weaker layers and removing false positives from the turbulence profile. 

}
The automated algorithms used by the Stereo-SCIDAR mean that the turbulence velocity profile is measured in real-time with no input from a user. The turbulence velocity profiles can then be fed into the turbulence strength profile reduction in real-time. This makes the Stereo-SCIDAR solution ideal for real-time support of existing and future astronomical adaptive optics systems. This data is immediately required for instrument modelling and development as well as for operational support as the next generation of sophisticated adaptive optics systems come on-line.

\comment{
\begin{itemize}
\item The low spatial resolution of the GFS model is not a significant problem for wind velocity profiling.
\item The radiosonde can be used to validate the Stereo-SCIDAR velocities, despite the spatial separation.
\item The wind velocity automatic turbulence velocity extraction algorithm functions satisfactorily.
\item{Wind velocity is the same as turbulence velocity}
\item{There is considerable more variability in wind direction than in wind speed. This is partially because wind direction identification of slow turbulent layers is inherently more difficult due to the fewer pixels covered by the covariance peak in the sample time, but it is also possible that gusting in the turbulence is more variable in direction than in magnitude.}
\end{itemize}
}
\newpage
\section*{Acknowledgements}
We are grateful to the Science and Technology Facilities Committee (STFC) for financial support (grant reference ST/J001236/1). FP7/2013-2016: The research leading to these results has received funding from the European Community's Seventh Framework Programme (FP7/2013-2016) under grant agreement number 312430 (OPTICON). The Isaac Newton Telescope is operated on the island of La Palma by the Isaac Newton Group in the Spanish Observatorio del Roque de los Muchachos of the Instituto de Astrofisica de Canarias. We are also grateful to the NOAA Earth System Research Laboratory for making the radiosonde and Global Forecast System (GFS) meteorological forecast data available on the web. MJT gratefully acknowledge support from the Science and Technology Facilities Council (STFC) in the form of a PhD studentship (ST/K501979/1). The Stereo-SCIDAR data used in this project is available upon request from the authors.
\bibliographystyle{mnras}
\bibliography{library}

\bsp	
\label{lastpage}
\end{document}